\documentclass[twocolumn,showpacs,preprintnumbers,amsmath,amssymb,floatfix]{revtex4}
\usepackage{graphicx}
\usepackage{epsfig}

\begin{document}

\title{Reaching the hydrodynamic regime in a Bose-Einstein condensate by suppression of avalanches}

\author{K.~M.~R. van der Stam}
\author{R.~Meppelink}
\author{J.~M. Vogels}
\author{P. van der Straten}\affiliation{Atom Optics and Ultrafast Dynamics, Utrecht University,\\
P.O. Box 80,000, 3508 TA Utrecht, The Netherlands}

\date{\today }

\begin{abstract} 
We report the realization of a Bose-Einstein condensate (BEC) in the hydrodynamic regime. The hydrodynamic regime is reached by evaporative cooling at a relative low density suppressing the effect of avalanches. With the suppression of avalanches a BEC containing 120$\cdot$10$^6$ atoms is produced. The collisional opacity can be tuned from the collisionless regime to a collisional opacity of more than 3 by compressing the trap after condensation. In the collisional opaque regime a significant heating of the cloud at time scales shorter than half of the radial trap period is measured. This is direct proof that the BEC is hydrodynamic.
\end{abstract}

\maketitle

The behavior of excitations in a BEC with energies larger than the mean field energy is determined by the mean free path of the atoms. Usually  the mean free path of the atoms is larger than the size of the sample (the collisionless regime). It would be of great interest to realize a BEC in the hydrodynamic regime, where the mean free path of the atom is less than the size of the condensate. In this situation the properties of the BEC are strongly influenced by the inter-atomic collisions. A hydrodynamic BEC would give the opportunity to investigate interesting properties of the condensate, for example, thermal excitations, heat conduction, shape oscillations, when there is only locally thermal equilibrium.

The transition from the collisionless to the hydrodynamic regime above the BEC transition temperature has been studied theoretically \cite{Khawaja} and experimentally \cite{Walraven}. For the situation below the BEC transition temperature theoretical discussions are given in Refs.~\cite{Stoof, Shenoy, Nikuni,Guery}. In this Letter we will discuss the experimental realization of a BEC in the hydrodynamic regime and study the excitations generated by three-body decay.

The most obvious way of reaching the hydrodynamic regime is creating a large and dense BEC. However, like shown in Ref. \cite{avalanches} the atom losses will increase strongly due to avalanches at such high densities, that are normally necessary for entering the hydrodynamic regime. This will severely limit the lifetime of the condensate in the hydrodynamic regime, as well as the collisional opacity. Therefore, Schuster \textit{et al.} \cite{avalanches} concluded that the collisionally opaque regime can hardly be reached in alkali BEC experiments. In the case of metastable helium BEC experiments to reach the hydrodynamic regime has so far been unsuccessful~\cite{HeAspect,HeLeduc}. A second way of realizing a hydrodynamic BEC is increasing the cross section for the elastic collisions by means of a Feshbach resonance. The increase in the cross section results in a large loss rate \cite{Feshbach}, which makes it an unsuitable approach for achieving a hydrodynamic BEC. Notable exceptions are the BEC's of molecules consisting of fermions tuned close to the unitarity limit~\cite{unitary}.
 
In this Letter we will show that it is possible to enter the hydrodynamic regime by following the first approach with a strong reduction of the effects of the problems mentioned above. This is done by evaporative cooling a cloud of atoms to the BEC temperature in a decompressed trap. The low trap frequencies lead to a relative low density of the cloud, which reduces the onset of avalanches. The suppression of avalanches results in our setup in a BEC containing 120$\cdot$10$^{6}$ sodium atoms at a density of 2.7$\cdot$10$^{14}$ atoms/cm$^{3}$. The low density suppresses the losses as a result of  avalanches, while the large physical size of the BEC results in a collisional opacity close to the hydrodynamic limit. The hydrodynamic regime can subsequently be entered deeply by compressing the trap after BEC is reached. The lifetime of the BEC in the hydrodynamic regime is more than 5 s, which is an increase with a factor of 25 compared to earlier work~\cite{avalanches}. 

During the evaporation towards the BEC transition the cloud will experience losses due to inelastic collisions between the atoms. These losses can be divided into three categories: one-, two- and three-body collisions, determined by the rate constants $G_{1,2,3}$. The one-body collisions are the collisions between the trapped atoms and the background gas. The pressure of our background gas (below $10^{-11}$ mbar, the limit of our detector) results in a lifetime of 260 s. These losses are dominant at densities below 5.9$\cdot$10$^{13}$ atoms/cm$^{3}$ or at temperatures above 1 $\mu K$. For reaching a large atom number BEC this is an important loss process; however, at the densities reached in our BEC it is no longer dominant.  The loss rate due to two-body collisions is given by $G_2 \cdot n$, with $G_{2}$ = 6$\cdot$10$^{-17}$ cm$^{3}$/s~\cite{G2} and $n$ is the density of the sample. This collision rate never exceeds the one- or three-body losses in our experiment and will therefore not be considered further in this Letter. 

The dominant loss process at the densities needed for the BEC transition in our case is three-body collisions. The loss rate for this process is given by $G_3 \cdot n^2$ with $G_{3}$ = 1.1$\cdot$10$^{-30}$ cm$^{6}$/s~\cite{G3}. The three-body collisions generate high energy atoms that will in the collisionless regime before they rethermalize be removed from the trap by evaporation. However, when the sample is in the hydrodynamic regime, the high energy atom will collide with other atoms before it leaves the trap resulting in an avalanche. In that case a large fraction of the energy generated by a three-body decay is released into the BEC resulting in a large loss rate of atoms and a strong heating of the sample. By contrast, in the collisionless regime only 3 atoms are lost without heating the sample. This striking difference in thermal properties makes the hydrodynamic regime very interesting, but extremely hard to reach. 
 
The apparatus is described elsewhere in more detail \cite{spinpol,grootBEC}. We start the experiment with a magneto-optical trap (MOT) containing 2.0$\cdot$10$^{10}$ sodium atoms. After spin polarizing in a high magnetic field 1.4$\cdot$10$^{10}$ atoms are loaded in a cloverleaf magnetic trap (MT). The magnetic field gradient of $118$ Gauss/cm in the radial direction and the curvature field in the axial direction of $42$ G/cm$^2$ lead to trap frequencies in the axial and radial direction of $\nu_{z}$ = 16 Hz and $\nu_{\rho}$ = 99 Hz at a field minimum of 3.4 G. We can cool the sample of atoms to degeneracy in 50 seconds.
 
\begin{figure}[t!]
\centering
\includegraphics[width=0.44\textwidth]{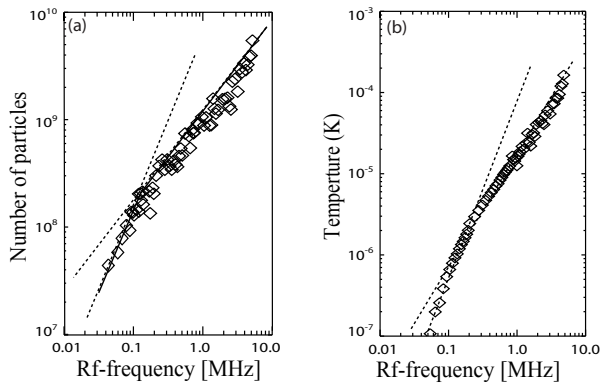}
\caption{Number of atoms (a) and the temperature (b) as a function of the rf-frequency for evaporative cooling with respect to the bottom of the trap. To make the decrease in the efficiency more clear two lines are drawn through the data.}
\label{RF}
\end{figure}

When we measure the number and the temperature of the particles during the evaporative cooling, the efficiency parameter $\alpha=\frac{\dot{T}/T}{\dot{N}/N}$ decreases at rf-frequencies below 200 kHz (Fig.~\ref{RF}). The density at this point in the evaporation is $(7\pm1)\cdot10^{13}$ atoms/cm$^{3}$, which is close to the density, where the three-body losses become the dominant loss process.
Despite the large loss rate we can still condense 35$\cdot$10$^{6}$ atoms at a density of 4.0$\cdot$10$^{14}$ atoms/cm$^{3}$. 

The collisional opacity $\kappa$ is given by $\kappa=\left\langle nr \right\rangle \sigma_s$, with $\sigma_s=8\pi a^2$ the s-wave cross section, $a$ the scattering length, and $\left\langle nr \right\rangle$ the average column density. Under the conditions of Fig.~\ref{RF} $\kappa$ = 1.3, which means that the mean free path of an atom in the condensate is less than its size making the BEC hydrodynamic. However, for a thorough investigation of the hydrodynamics of the BEC it is necessary to enter the hydrodynamic regime even deeper. Therefore, a BEC containing more atoms is needed. This is realized by suppressing the losses due to three-body collisions during the evaporation. Since the three-body losses scale quadratically with density, this suppression is achieved by reducing the density. In order to achieve this, the trap is decompressed after 42 seconds of evaporation, which corresponds with an rf-frequency of 2.04 MHz with respect to the bottom of the trap. The decompression is done adiabatically in 2 s. Before this point three-body losses are not the limiting loss process because the density is always below 2$\cdot$10$^{13}$ atoms/cm$^{3}$ yielding a loss rate of 4$\cdot$10$^{-4}$  s$^{-1}$. For the decompression the trap frequencies can be lowered in either axial or radial direction. 

\begin{figure}[t!]
\centering
\includegraphics[width=0.44\textwidth]{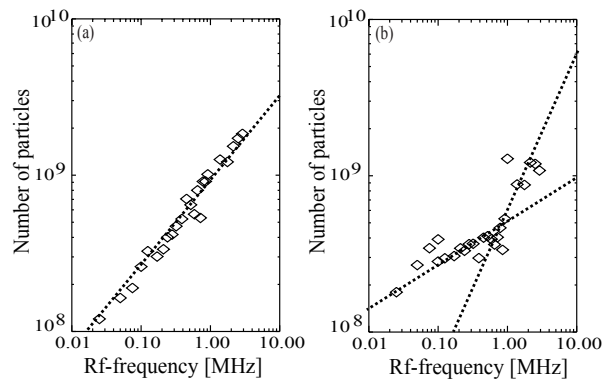}
\caption{The number of particles during the evaporation in a radial decompressed trap (a) and in an axial decompressed trap (b).}
\label{decomp}
\end{figure}

In Fig.~\ref{decomp}a the number of particles during the evaporative cooling is given for a radial decompressed trap. The radial trap frequency is in this case lowered from 99 Hz to 50 Hz resulting in an average trap frequency $\bar{\omega}\equiv\sqrt[3]{\omega_r\omega_r\omega_z}$ = 34 Hz. Combined with the results of the temperature as a function of rf-frequency, we observe no decrease in efficiency. The evaporation in this trap results in a BEC containing 60$\cdot$10$^{6}$ atoms at a density of 2.7$\cdot$10$^{14}$ atoms/cm$^{3}$. 
When the decompression takes place in the axial direction (16 Hz $\rightarrow$ 4 Hz) instead of the radial direction the efficiency is even higher (Fig.~\ref{decomp}b). Note, that this decompression results in the same average trap frequency $\bar{\omega}$ as above and thus the same decrease in density. The evaporation in the axial decompressed trap leads to 120$\cdot$10$^{6}$ atoms in the BEC. This is as far as we know the largest condensate starting from an optical trap. Note, that we loose less than a factor of 120 in the number of atoms during the evaporation indicating an efficient suppression of the losses. The increase in the efficiency parameter $\alpha$ indicated in Fig.~\ref{decomp}b by the dotted lines suggests that an earlier decompression would even lead to more condensed atoms. However, by carefully optimizing the sequence it turned out that the decompression after 42 s of evaporation leads to the largest number of atoms.

\begin{figure}[t!]
\centering
\includegraphics[width=0.31\textwidth]{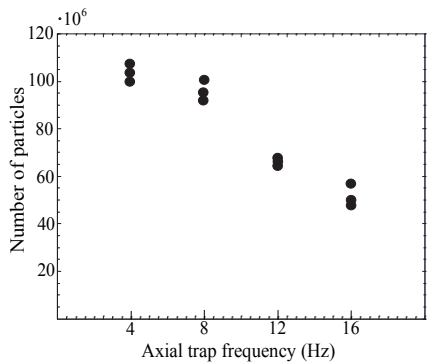}
\caption{The number of particles in the BEC as a function of the axial trap frequency with $\bar{\omega}$ = 34 Hz.}
\label{decomp2}
\end{figure}
In Fig.~\ref{decomp2} the results of a more quantitative analysis of decompression are given. In this figure the number of atoms in the BEC is plotted for different decompression scenarios. The radial trap frequency is adjusted in such way that for every measurement the average trap frequency is identical ($\bar{\omega}$ = 34 Hz). The decompression discussed above in the axial and radial direction represent the two extremes in this figure. The number of particles shows a monotone increasing behavior with decreasing axial trap frequency. In all cases, the initial temperature, density and therefore, three-body collision rate are identical. The only difference is the aspect ratio of the cloud. In the radial decompressed trap the collisional opacity $\kappa$ is larger, so that atoms undergoing three-body collisions are more likely to generate avalanches. Furthermore, a large $\kappa$ can hamper the evaporative cooling. Either way, these measurements are a clear sign that we have reached the hydrodynamic regime. In the axial decompressed trap the avalanches are suppressed, until BEC is reached. At the BEC transition the collisional opacity has increased to 0.8 resulting in a  BEC at the cross-over between the collisionless and the hydrodynamic regime. Since the suppression of avalanches in this case is already effective at 8 Hz, further decompression to 4 Hz does not lead to a strong increase in the number of particles in the BEC.


Above we have described how a large atom number BEC with $\kappa$ close to 1 is realized by the suppression of avalanches. After BEC transition is reached $\kappa$ can be tuned from the collisionless regime to a value of 3.2 by compressing in 100 ms the trap in axial direction. In the remainder of this Letter the adjustability of $\kappa$ is used for two kind of experiments. First, the influence of the avalanches on the loss rate in investigated. Second, a direct proof of the hydrodynamics is presented for $\kappa$ = 3.2 by studying the heating due to avalanches. 

\begin{figure}[t!]
	\centering
		\includegraphics[width=0.37\textwidth]{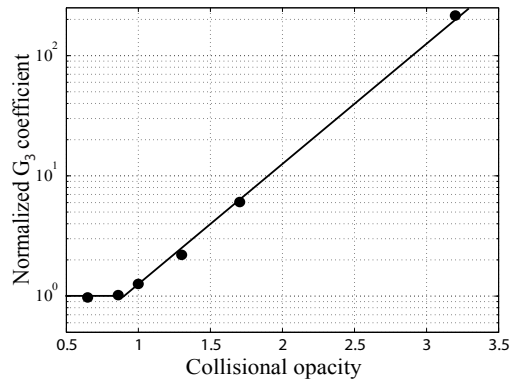}
	\caption{The increase in $G_3$ coefficient as a function of the collisional opacity $\kappa$. The determined $G_3$ coefficient is divided by the $G_3$ coefficient in the collisionless regime \cite{G3}. The solid line is a guide to the eye. }
	\label{gainG3}
\end{figure}

Previous work \cite{avalanches} has shown that the lifetime of a BEC in the hydrodynamic regime is severely limited by avalanches. In this experiment the avalanches increases the $G_3$ coefficient by a factor of 8. We have measured the $G_3$ coefficient as a function of $\kappa$ (Fig.~\ref{gainG3}). For each value of $\kappa$ the $G_3$ coefficient is determined from the lifetime of the BEC. As seen from Fig.~\ref{gainG3}, the $G_3$ coefficient increases with more than two orders of magnitude by increasing $\kappa$. At collisional opacities below 0.8 the loss rate is determined by the $G_3$ coefficient in the collisionless regime, indicating that there are no avalanches. This is slightly above the predicted value of 0.69 for the onset of the avalanches \cite{avalanches}. This small discrepancy was already mentioned by Streed \textit{et al.} \cite{avalanchesKetterle}. The difference between the predicted value and the measurements indicates that the transition from the collisionless regime to the hydrodynamic regime is not as sharp as expected. The energy of a three-body decay is in the order of 3000~$T_c$, so it is expected that the $G_3$ coefficient will increase even further with increasing $\kappa$. In our experiment we are not able to compress the MT further or to condense more atoms, so we could not determine the upper limit of $G_3$.
   
When the trap is fully compressed after reaching the BEC transition $\kappa$ = 3.2 is reached at a density of 6.6$\cdot$10$^{14}$ atoms/cm$^{3}$. The increased density due to the compression and the effect of the avalanches results in a reduction of the lifetime to below 20 ms. The lifetime is sufficient for several experiments and can be increased by tuning the BEC to a slightly lower collisional opacity. 

\begin{figure}[t!]
\centering
\includegraphics[width=0.36\textwidth]{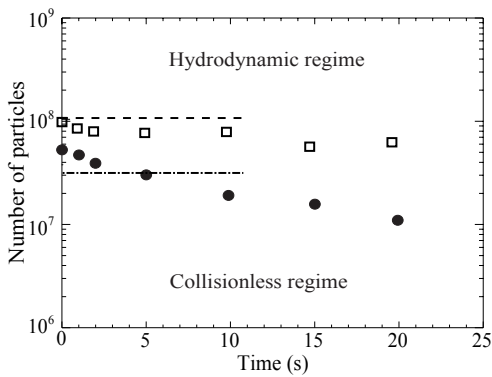}
\caption{The number of particles in the BEC as a function of the storage time for a hydrodynamic BEC with $\kappa$ = 1.7 (circles), and a BEC in the collisionless regime with $\kappa$ = 0.8 (squares). The horizontal lines indicates $\kappa$ = 1.}
\label{lifetime}
\end{figure}

In Fig.~\ref{lifetime} the lifetime of the BEC in the hydrodynamic regime is studied in more detail by measuring the number of particles in the BEC as a function of the storage time for a hydrodynamic BEC with $\kappa$ = 1.7 (circles), and a BEC in the collisionless regime with $\kappa$ = 0.8 (squares). During the storage time an rf-shield is applied at 100 kHz generating a trap depth of 5 $\mu$K to avoid strong heating. The lifetime of the condensate in the hydrodynamic and collisionless regime is 10 and 20 s, respectively. The horizontal lines indicate the number of particles needed for $\kappa$ = 1.0. Note, that the crossover between the hydrodynamic and collisionless regime is different for the two situations, because the trap frequencies are different. We can see that the hydrodynamic BEC stays in the hydrodynamic regime for more than 5 s, which is an increase with more than a factor of 25 compared to earlier work \cite{avalanches}. The hydrodynamic BEC was produced with radial decompression during the evaporation, proving that we can evaporative cool in the hydrodynamic regime. Since the losses are larger during the cooling the BEC contains less atoms. Note, that the decay of the hydrodynamic BEC is much faster at small time scales indicating avalanche enhanced losses. 

\begin{figure}[t!]
	\centering
		\includegraphics[width=0.36\textwidth]{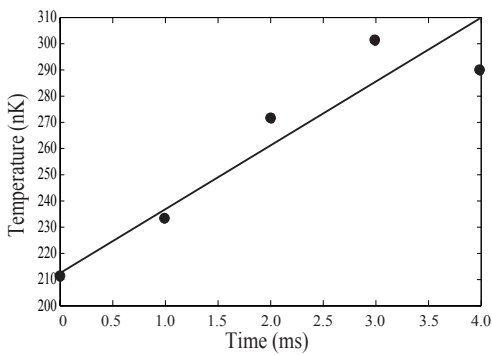}
	\caption{The heating of the sample for a strongly hydrodynamic BEC with $\kappa$ = 3.2.  The solid line is a guide to the eye.}
	\label{heating}
\end{figure}

The collisional opacity of 3.2 results in a s-wave collision probability of 0.96. This means that an atom has chance of 96 \% to collide with another atom, while traveling from the center to the edge of the condensate. In this situation a significant amount of the energy generated in a three-body collision is distributed over the BEC. In the collisionless regime the high energy atom will be evaporated by the rf-shield before the energy is released into the BEC. Therefore, in the hydrodynamic regime heating can occur on a time scale faster than half the radial trap period, which is 5 ms in our case. When we measure the temperature of the strongly hydrodynamic sample as a function of the storage time we see an increase of the temperature due to heating of more than 35\% within 4 ms (Fig.~\ref{heating}). 
This is a direct  proof that the BEC is deeply in the hydrodynamic regime. Even when an rf-shield is applied this heating is not significantly suppressed. 
 
In this Letter we have described the experimental realization of a hydrodynamic BEC. This is achieved by suppressing the avalanches during the evaporation towards the BEC transition resulting in a BEC containing 120$\cdot$10$^{6}$ sodium atoms with a collisional opacity of 0.8. The collisional opacity is low enough to suppress the avalanches during generation of the BEC, while it is high enough to yield the possibility to enter the hydrodynamic regime deeply by compressing the trap, after the BEC transition is reached. The collisional opacity can be tuned from the collisionless  to the hydrodynamic regime with a collisional opacity of 3.2 by axial compression. Furthermore, the lifetime of the BEC in the hydrodynamic regime is more than 5 s, which makes it a very good starting point for further research on the hydrodynamic behavior of the BEC.

\end{document}